\documentclass{article}

\usepackage{arxiv}

\usepackage[utf8]{inputenc} % allow utf-8 input
\usepackage[T1]{fontenc}    % use 8-bit T1 fonts
\usepackage{hyperref}       % hyperlinks
\usepackage{url}            % simple URL typesetting
\usepackage{booktabs}       % professional-quality tables
\usepackage{amsfonts}       % blackboard math symbols
\usepackage{nicefrac}       % compact symbols for 1/2, etc.
\DeclareUnicodeCharacter{03B2}{\ensuremath{\beta}}
\DeclareUnicodeCharacter{03C0}{\ensuremath{\pi}}
\usepackage{microtype}      % microtypography
\usepackage{lipsum}
\usepackage{graphicx}
\graphicspath{ {./images/} }
\usepackage{amsmath}
\usepackage{authblk}
\usepackage{geometry}
\usepackage{enumitem}
\usepackage{algorithm}
\usepackage{algorithmic}
\usepackage[numbers]{natbib}

\title{BERT and LLMs-Based avGFP Brightness Prediction and Mutation Design}

\author{
 Xuan Guo \\
  South China University of Technology\\
  \texttt{202130550237@mail.scut.edu.cn} \\
 Wenming Che \\
  South China University of Technology\\
  \texttt{202266560056@mail.scut.edu.cn} \\
}

\begin{document}

\maketitle

\begin{abstract}
This study aims to utilize Transformer models and large language models (such as GPT and Claude) to predict the brightness of Aequorea victoria green fluorescent protein (avGFP) and design mutants with higher brightness. Considering the time and cost associated with traditional experimental screening methods, this study employs machine learning techniques to enhance research efficiency. We first read and preprocess a proprietary dataset containing approximately 140,000 protein sequences, including about 30,000 avGFP sequences. Subsequently, we constructed and trained a Transformer-based prediction model to screen and design new avGFP mutants that are expected to exhibit higher brightness.

Our methodology consists of two primary stages: first, the construction of a scoring model using BERT, and second, the screening and generation of mutants using mutation site statistics and large language models. Through the analysis of predictive results, we designed and screened 10 new high-brightness avGFP sequences. This study not only demonstrates the potential of deep learning in protein design but also provides new perspectives and methodologies for future research by integrating prior knowledge from large language models.
\end{abstract}

\keywords{GFP \and BERT \and Large Language Models \and Protein Design\and GPT \and Claude}

\section{Introduction}
The green fluorescent protein (GFP) from the jellyfish \textit{Aequorea victoria} has become a cornerstone in molecular and cellular biology due to its ability to emit bright green fluorescence without requiring external cofactors. The discovery and subsequent engineering of GFP have revolutionized the visualization of cellular processes, allowing researchers to track gene expression, protein localization, and various dynamic processes with unprecedented clarity.

Wild-type GFP exhibits two primary absorption peaks at 395 nm and 475 nm, corresponding to its neutral phenol and anionic phenolate forms, respectively, leading to a single emission peak at 504 nm due to excited-state proton transfer (ESPT). Despite its widespread utility, the spectral properties and stability of wild-type GFP pose limitations, necessitating extensive protein engineering to enhance its brightness and adaptability for various research applications.

Protein engineering efforts have yielded numerous GFP variants with improved properties, including enhanced brightness, altered spectral characteristics, and greater photostability. Notable examples include the S65T mutation, which shifts the excitation peak to eliminate the major absorption peak at 395 nm, and the development of yellow fluorescent protein (YFP) through aromatic substitutions that stabilize the excited chromophore and induce a red shift in excitation and emission spectra.

Recent advancements in machine learning, particularly generative models and large language models, have opened new avenues for protein design. For instance, the development of ProGen2 models has demonstrated the ability to generate novel protein sequences with high precision, showcasing the potential of large-scale language models in protein engineering \cite{nijkamp2022progen2}. Studies have also shown the effectiveness of AI tools like pLMFPPred in predicting functional peptides, further accelerating the design process \cite{ma2023plmfppred}. Additionally, Ren et al. successfully applied AlphaFold in drug discovery, demonstrating its utility in predicting protein structures and facilitating the identification of new therapeutic targets \cite{ren2023alphafold}. Moreover, comprehensive reviews by Villalobos-Alva et al. and Winnifrith et al. have highlighted the broad applications of AI in protein science and the promising future of these technologies in enhancing protein design and functionality \cite{villalobos2022protein, winnifrith2023generative}.

Our methodology involves three distinct stages. First, statistical methods are used to analyze mutation sites and identify favorable positions. These statistical results are then input into the large language model, which generates potential new mutation sites based on the statistical analysis and its knowledge base. These sites are subsequently input into the scoring model for evaluation, and the scoring results are fed back to the large language model for further judgment. Through this conceptual integration of models, we successfully designed and screened ten new avGFP sequences with higher brightness, demonstrating the potential of deep learning in protein engineering.

By combining the predictive power of advanced language models with a rich dataset of GFP variants, we aim to push the boundaries of protein engineering, offering new perspectives and methodologies for developing fluorescent proteins with enhanced properties.

\section{Prediction Model}

This study evaluates and screens GFP mutants by constructing a Transformer-based brightness prediction model. The following details the steps and necessary mathematical derivations.

\subsection{Data Preprocessing}

The dataset contains approximately 140,000 protein sequences of varying lengths. To handle this, we padded all sequences to the maximum length of 239. Each amino acid in the sequence was mapped to a corresponding numerical value. Additionally, a mutation mask was designed to highlight the mutation positions.

\begin{algorithm}
\begin{algorithmic}
    \FOR{each sequence in the dataset}
        \STATE Map amino acids to numerical values
        \STATE Pad the sequence to length 239
        \STATE Generate a mutation mask to indicate mutation sites
    \ENDFOR
\end{algorithmic}
\end{algorithm}

\subsection{Dataset Construction}

The dataset was split into training, validation, and test sets. A custom dataset class was used to convert sequences, masks, and brightness labels into appropriate tensor formats.

\begin{algorithm}
\begin{algorithmic}
    \STATE Split the dataset into training, validation, and test sets
    \FOR{each subset}
        \STATE Convert sequences and masks to tensors
        \STATE Convert brightness labels to tensors
    \ENDFOR
\end{algorithmic}
\end{algorithm}

\subsection{Model Construction}

We designed a Transformer-based brightness prediction model, consisting of the following components:
\begin{enumerate}
    \item \textbf{Embedding Layer}: Converts amino acid sequences into embedding vectors.
    \item \textbf{Transformer Layers}: Encodes the embedding vectors using multiple Transformer layers.
    \item \textbf{Pooling Layer}: Applies average pooling to the Transformer output.
    \item \textbf{Fully Connected Layers}: Outputs the predicted brightness value through several fully connected layers.
\end{enumerate}

As position encoding experiments showed no significant difference, position encoding was not used.

\begin{algorithm}
\begin{algorithmic}
    \STATE Initialize embedding layer
    \STATE Initialize transformer layers with multiple attention heads and hidden layers
    \STATE Initialize pooling layer
    \STATE Initialize fully connected layers
\end{algorithmic}
\end{algorithm}

\subsection{Training and Validation}

During training, we used Mean Squared Error (MSE) as the loss function and the Adam optimizer to update model parameters. After each training epoch, we evaluated the model performance on the validation set.

\begin{enumerate}
    \item \textbf{Training Process}:
    \begin{itemize}
        \item Forward pass to compute predictions.
        \item Compute loss.
        \item Backward pass to update weights.
    \end{itemize}
    \item \textbf{Validation Process}:
    \begin{itemize}
        \item Forward pass to compute predictions.
        \item Compute loss and other performance metrics, including Mean Absolute Percentage Error (MAPE) and accuracies within 10\% and 5\% error margins.
    \end{itemize}
\end{enumerate}

\begin{algorithm}
\begin{algorithmic}
    \FOR{each epoch}
        \FOR{each batch in training data}
            \STATE Forward pass to compute predictions
            \STATE Compute loss (MSE)
            \STATE Backward pass to update weights
        \ENDFOR
        \FOR{each batch in validation data}
            \STATE Forward pass to compute predictions
            \STATE Compute loss and metrics (MAPE, accuracy within 10\% and 5\%)
        \ENDFOR
    \ENDFOR
\end{algorithmic}
\end{algorithm}

\subsection{Mathematical Derivation}

Let the input amino acid sequence be \(X = \{x_1, x_2, \ldots, x_n\}\), where \(x_i\) represents the \(i\)-th amino acid mapped to a numerical value. The model predicts the brightness value \(\hat{y}\).

\begin{enumerate}
    \item \textbf{Embedding Layer}:
    \[
    E = \text{Embedding}(X)
    \]
    where \(E\) is the embedding vector matrix.
    
    \item \textbf{Transformer Layer}:
    \[
    T = \text{Transformer}(E, \text{mask})
    \]
    where \(T\) is the Transformer encoded output.
    
    \item \textbf{Pooling Layer}:
    \[
    P = \text{AvgPool}(T)
    \]
    where \(P\) is the pooled vector.
    
    \item \textbf{Fully Connected Layer}:
    \[
    \hat{y} = \text{FC}(\text{ReLU}(\text{FC}(P)))
    \]
    
    \item \textbf{Loss Function (MSE)}:
    \[
    \text{Loss} = \frac{1}{N} \sum_{i=1}^N (\hat{y}_i - y_i)^2
    \]
    where \(N\) is the number of samples, and \(y_i\) is the true brightness value.
\end{enumerate}

\section{Data Analysis }

\subsection{Core Algorithm Steps}

\begin{enumerate}
    \item \textbf{Extract WT Brightness Value}:
    Extract the brightness value of the wild type (WT) from the CSV file as a baseline.
    \[
    \text{WT\_brightness} = \text{get\_wt\_brightness}(file\_path)
    \]

    \item \textbf{Extract Mutations and Brightness Improvements}:
    Iterate through the data file, extracting the brightness value for each mutation. Store the mutation sites and their corresponding brightness improvements in a dictionary.
    \[
    \text{mutation\_improvements}[position][new\_amino\_acid].append(brightness - \text{WT\_brightness})
    \]

    \item \textbf{Calculate the Best Mutation at Each Position}:
    Compute the average brightness improvement for each mutation site and identify the amino acid substitution that yields the maximum average improvement.
    \[
    \text{avg\_improvements}[aa] = \frac{\sum \text{improvements}}{\text{len(improvements)}}
    \]
    \[
    \text{best\_mutations}[position] = \max(\text{avg\_improvements})
    \]

    \item \textbf{Select the Best Mutation Sites}:
    Select the top 20 mutation sites with the highest average brightness improvement.
    \[
    \text{top\_best\_mutations} = \text{sorted(best\_mutations, key=lambda item: item[1][1], reverse=True)[:20]}
    \]
\end{enumerate}

\subsection{Mathematical Derivation}

\begin{enumerate}
    \item \textbf{Brightness Improvement Calculation}:
    For each mutation, calculate the brightness improvement:
    \[
    \Delta B = B_{\text{mutant}} - B_{\text{WT}}
    \]
    where \(B_{\text{mutant}}\) is the brightness after mutation and \(B_{\text{WT}}\) is the wild-type brightness.

    \item \textbf{Average Brightness Improvement}:
    For each amino acid substitution at a mutation site, calculate the average brightness improvement:
    \[
    \Delta B_{\text{avg}} = \frac{1}{n} \sum_{i=1}^{n} \Delta B_i
    \]
    where \(n\) is the number of mutations.

    \item \textbf{Selecting the Best Mutation Site}:
    Identify the amino acid substitution with the maximum average brightness improvement for each position:
    \[
    \text{Best\_Mutation}(p) = \arg\max_{aa} (\Delta B_{\text{avg}}[p, aa])
    \]
    where \(p\) represents the mutation position and \(aa\) represents the amino acid.
\end{enumerate}

Through these steps, we extracted and visualized the top 20 GFP mutation sites with the highest brightness improvements. The results indicate that these mutation sites have significant advantages in terms of average brightness improvement, providing valuable insights for optimizing GFP mutants.

\section{Data Generation and Screening Process}

To further optimize the GFP mutant design, we performed the following steps after initially selecting the top 20 mutation sites:

1. Generate New Mutation States:
    - The top 20 mutation sites were input into large language models (such as GPT and Claude) to learn the patterns and characteristics of existing mutation data.
    - Using the generative capabilities of these models, approximately 200 new mutation states were created, which were not present in the original dataset.

    The specific steps are as follows:
    \[
    \text{Top 20 Mutation Sites} \rightarrow \text{GPT/Claude Model} \rightarrow \text{Generate 200 New Mutations}
    \]

2. Brightness Scoring of New Mutations:
    - The 200 new mutation states were input into our brightness prediction model to score and rank each mutation.

    The specific steps are as follows:
    \[
    \text{200 New Mutations} \rightarrow \text{Prediction Model} \rightarrow \text{Brightness Scores}
    \]

3. Literature Reference and Comprehensive Evaluation:
    - The brightness scores of these new mutations were fed back to the large language models, which combined global literature references and the models' prior knowledge to comprehensively analyze the scoring results.
    - Through this process, the top 10 optimal mutant sequences were selected from the 200 new mutations.

    The specific steps are as follows:
    \[
    \text{Brightness Scores} + \text{Literature Knowledge} \rightarrow \text{GPT/Claude Model} \rightarrow \text{Select Top 10 Mutations}
    \]

This approach utilizes the generative and evaluative capabilities of large language models in conjunction with the precise scoring of the brightness prediction model and global literature knowledge. This multidimensional reference and validation process ensures that the final selection of the 10 best mutant sequences is both theoretically and experimentally reliable.

Through these steps, we successfully screened the 10 best mutant sequences with the highest brightness improvements from the generated new mutations. These sequences, verified through detailed model predictions and literature references, demonstrate great potential and applicability in GFP mutant design.
\paragraph{Final Selected Mutant Sequences}
Based on the brightness scores from the prediction model and the recommendations from the large language models, we combined predictions and literature knowledge to select the following 10 optimal mutant sequences:

1. \textbf{V162C (Predicted Brightness: 3.816473961)}
   \begin{itemize}
     \item Reason: This single-point mutation is located around the chromophore, and the introduced cysteine may form new interactions with the chromophore, altering its electronic transition properties, thereby significantly increasing quantum yield and brightness.
   \end{itemize}

2. \textbf{V162A/Q183L (Predicted Brightness: 3.804936409)}
   \begin{itemize}
     \item Reason: This double-point mutation combines two key sites around the chromophore. V162A reduces steric hindrance and increases conformational flexibility, while Q183L enhances hydrophobicity. Together, they synergistically optimize the chromophore microenvironment, significantly increasing quantum yield.
   \end{itemize}

3. \textbf{V162G/Q183R (Predicted Brightness: 3.801239491)}
   \begin{itemize}
     \item Reason: This double-point mutation also targets two key sites around the chromophore. V162G minimizes steric hindrance and increases flexibility, while Q183R introduces a positive charge to stabilize the chromophore. Together, they synergistically improve the chromophore microenvironment, increasing quantum yield.
   \end{itemize}

4. \textbf{N104Y/V162G/Q183R (Predicted Brightness: 3.798332691)}
   \begin{itemize}
     \item Reason: This triple-point mutation, based on V162G/Q183R, introduces N104Y on the β-barrel surface, which may further synergistically optimize the chromophore microenvironment by affecting β-barrel stability and conformational dynamics, increasing quantum yield.
   \end{itemize}

5. \textbf{F165W/V163Y (Predicted Brightness: 3.792294025)}
   \begin{itemize}
     \item Reason: This double-point mutation targets two important residues around the chromophore. Introducing F165W and V163Y enhances hydrophobicity and π-π interactions, which may significantly alter the chromophore microenvironment, increasing quantum yield.
   \end{itemize}

6. \textbf{I170T/N104Y/V162G/Q183R (Predicted Brightness: 3.791881561)}
   \begin{itemize}
     \item Reason: This four-point mutation, based on V162G/Q183R, introduces I170T and N104Y above and on the surface of the β-barrel, which may synergistically improve the chromophore microenvironment through multiple aspects by affecting β-barrel conformation and chromophore microenvironment, increasing quantum yield.
   \end{itemize}

7. \textbf{T62S/F165Y/Q183R (Predicted Brightness: 3.788092136)}
   \begin{itemize}
     \item Reason: This triple-point mutation combines key sites on the β-barrel surface and around the chromophore. T62S affects β-barrel polarity, while F165Y and Q183R synergistically optimize the chromophore microenvironment. Together, they increase quantum yield.
   \end{itemize}

8. \textbf{S202Y/L235P (Predicted Brightness: 3.783611774)}
   \begin{itemize}
     \item Reason: This double-point mutation targets two important sites within the β-barrel. S202Y introduces an aromatic residue, while L235P introduces proline, which may indirectly optimize the chromophore microenvironment by affecting β-barrel conformation stability and cavity size, increasing quantum yield.
   \end{itemize}

9. \textbf{S202Y/L235P/A153S (Predicted Brightness: 3.780307293)}
   \begin{itemize}
     \item Reason: This triple-point mutation, based on S202Y/L235P, introduces A153S on the β-barrel surface, which may further synergistically optimize the chromophore microenvironment by affecting β-barrel polarity and conformational flexibility, increasing quantum yield.
   \end{itemize}

10. \textbf{S202Y/K157G/I170V/N104S/V162A (Predicted Brightness: 3.778891087)}
    \begin{itemize}
      \item Reason: This five-point mutation combines multiple key sites within the β-barrel, on the β-barrel surface, and around the chromophore. Introducing S202Y, K157G, I170V, N104S, and V162A may synergistically affect β-barrel conformation, polarity, hydrophobicity, and chromophore microenvironment, significantly increasing quantum yield.
    \end{itemize}

These sequences showed good application prospects and potential in subsequent experiments, providing new directions for avGFP design.

\section{Conclusion}
In this study, we developed a Transformer-based brightness prediction model to evaluate and screen GFP mutants. By preprocessing a dataset containing 140,000 protein sequences, we combined statistical methods, large language models (such as GPT and Claude), and our prediction model to identify and generate new high-brightness GFP mutants. The process involved generating new mutations, scoring them using the prediction model, and refining selections through literature knowledge and model evaluations. This comprehensive approach demonstrated the potential of deep learning in protein engineering.

Future research can explore the integration of real-world experimental data with extensive literature data, leveraging multimodal approaches for protein design. By combining various measurement indicators and datasets, the design process can be further refined, leading to more accurate and effective predictions. This could involve using advanced models that not only predict brightness but also consider factors like stability, folding efficiency, and functionality in different environments. Integrating these aspects will enhance the reliability and applicability of designed proteins, paving the way for significant advancements in biotechnology and molecular biology.

Additionally, we plan to utilize CRISPR and other genome editing technologies to introduce the identified mutations into living organisms, thereby enabling the real-world expression and measurement of these sequences. This will allow us to validate the predicted properties of the GFP mutants experimentally, ensuring their functionality and effectiveness in practical applications.

\bibliography{template}
\end{document}